\documentclass[a4paper, 12pt]{article}
\usepackage{graphicx}
\usepackage{amsmath}
\usepackage{amssymb}
\def \ni {\noindent}
\def \vs  {\vskip5mm}
\def \be {\begin{equation}}
\def \ee {\end{equation}}
\def \bea {\begin{eqnarray}}  
\def \eea {\end{eqnarray}}  
\def \mea {\nonumber\\}
\def \R {{\widehat \rho}}
\def \A {{\widehat A}}
\def \B {{\widehat B}}
\def \H {{\widehat H}}
\def \W {{\mathcal W}}
\def \half {{\textstyle \frac{1}{2}}}

\def\psistate {{|\psi\rangle}}

\def \Del {{\widehat \Delta}}
\def \Hil {{\mathcal H}}

\def \qO {{\widehat q}\,}
\def \pO {{\widehat p}\,}

\def \Par {{\widehat \Pi}}
\def\leftder {\stackrel{\leftarrow}{\partial}}
\def\rightder{\stackrel{\rightarrow}{\partial}}
\def \qBL {{q_{BL}}}
\def \pBL {{p_{BL}}}
\def \qBR {{q_{BR}}}
\def \pBR {{p_{BR}}}
\def \subspace {{\cal S}_{\varphi_0}}
\begin{document}    
\begin{titlepage}     
     
\title{The quantum state vector in phase space and Gabor's windowed Fourier transform}     
     
\author{A.J. Bracken\footnote{{\em Email:} a.bracken@uq.edu.au} and P. Watson\footnote{{\em Email:} pw.cmp@optusnet.com.au}\\Centre for Mathematical Physics\\Department of Mathematics\\    
University of Queensland\\Brisbane 4072\\Queensland\\Australia}     
     
\date{}     
\maketitle     
     
\begin{abstract}
\ni
Representations of quantum state vectors by complex  phase space amplitudes, complementing the description of the density operator 
by the Wigner function, have been defined by applying the Weyl-Wigner transform to dyadic operators, 
linear in the state vector and  anti-linear in a fixed `window state vector'.  
Here aspects of this construction are explored, and a connection is established with 
Gabor's `windowed Fourier transform'.  
The amplitudes that arise for simple quantum states from various choices of window are presented as illustrations. 
Generalized Bargmann representations of the state vector appear as  special cases, associated with  Gaussian windows.    
For every choice of window, amplitudes lie in a corresponding linear subspace of square-integrable functions on phase space.  
A generalized Born interpretation of amplitudes is described, with both the  Wigner function and a generalized Husimi function 
appearing as quantities linear in an amplitude and anti-linear in its complex conjugate. Schr\"odinger's time-dependent and 
time-independent equations are represented on phase space amplitudes, and their solutions described in simple cases.     
    
\end{abstract}

\ni
\end{titlepage}

\section{Introduction}

\vs\ni
The phase space formulation of quantum mechanics \cite{dubin,folland,zachos,gosson2}
now plays a central role in 
theoretical quantum optics \cite{schleich} and quantum tomography \cite{leonhardt}, and it has also become an important vehicle for investigations of 
fundamental questions such as 
the nature of quantization \cite{flato,kontsevich} and the quantum-classical interface \cite{littlejohn,bracken1,bracken2}.  
New features continue to be revealed by ongoing 
investigations of the underlying structures \cite{bracken3,bracken4,cassinelli}.
\vs\ni
The overwhelming majority of the very large number of studies in this field 
have focussed on the Wigner function $W$, the real-valued quasiprobability density on phase space $\Gamma$ which is  
the representative of the density operator
${\widehat \rho}$ under the action of the Weyl-Wigner transform $\W$. 
However, in the last two decades a small number of studies \cite{torres,harriman,wlodarz,ban,oliveira,gosson,smith1} 
have gradually made clear how the state vector 
$|\psi\rangle$, which is after all a 
more fundamental object than ${\widehat \rho}$, can also be depicted in the phase space picture, 
as a complex-valued amplitude $\Psi$ on $\Gamma$. The recent work of Smith \cite{smith1} in particular 
shows that this is achieved most simply by applying
$\W$ to a suitable multiple of the dyadic operator $|\psi\rangle\langle \varphi_0|$, 
where $|\varphi_0\rangle$ is an 
arbitrary normalized state vector that, once chosen, is held fixed while $|\psi\rangle$ 
ranges over all states of the given quantum system.   From this definition 
it follows that $\Psi$ is a multiple of the 
two-state Wigner function \cite{hillery} $W_{\psi \varphi_0}$, which is thereby 
given an important role in the phase space formulation.

\vs\ni
Several questions concerning this important extension of the phase   space formulation 
suggest themselves:
\begin{itemize}
\item
In what sense is $|\psi\rangle\langle\varphi_0|$, and hence its image $\Psi$, equivalent to $|\psi\rangle$ 
in  quantum mechanical calculations? 
\item
What relationships do the Wigner function $W$ and  the Husimi function
\cite{husimi} have
with the phase space amplitude $\Psi$?  
\item 
What form does the amplitude
$\Psi$ take for special system states $|\psi\rangle$ such as  coherent states, or eigenstates of position or momentum, for 
different choices of $|\varphi_0\rangle$?  
\item For a given $|\psi\rangle$, how does
the choice of $|\varphi_0\rangle$ influence the structure of $\Psi$ 
and its place in the set of functions on $\Gamma$? 
\item What forms do the time-dependent and time-independent 
Schr\"odinger equations take on phase space amplitudes,
and how do their solutions look in simple cases, for various choices of $|\varphi_0\rangle$?
\item What is the nature of the mapping from 
$|\psi\rangle$ to $\Psi$, when viewed as a transform of the coordinate space wavefunction $\psi$,   
and how does it relate to other, well-known
transforms in the literature? 
\end{itemize}

\vs\ni
These are  the questions that we attempt to address in what follows.  With regard to the  last question, 
we shall see that the transform in question 
is closely related to Gabor's `windowed Fourier transform' \cite{gabor}, widely used 
in the signal processing literature \cite{cohen1,cohen2} and also called 
there the `short-time Fourier transform'. 
Accordingly we shall refer to the fixed state $|\varphi_0\rangle$ that appears 
in the definition of 
phase space amplitudes $\Psi$ as the window vector, 
window state, or simply `the window' in what follows.  Other names for closely related objects in the 
literature  are `probe functions' \cite{torres} and `drone states' or `fiducial states' \cite{smith1}.
We shall see also that the choice of an oscillator ground state as window gives a $\Psi$ that is, up to a 
factor independent of the state $|\psi\rangle$, the well-known Bargmann wavefunction \cite{bargmann}, 
which is thereby seen as a precursor to the more recent efforts 
\cite{torres,harriman,wlodarz,ban,oliveira,gosson,smith1}  
to represent the state vector in phase space.

\vs\ni
For simplicity of presentation in what follows, we consider mainly one linear degree of freedom, 
ignore spin degrees of freedom, and treat all variables as dimensionless, setting Planck's constant $\hbar$ equal to $1$.
We use hats to label operators on the usual complex Hilbert space $\Hil$ of quantum mechanics. 
Variables without hats are defined on $\Gamma$, unless otherwise specified.

\section{Definition of phase space amplitudes} 
\vs\ni
Our starting point is the observation that  pure
state vectors $|\psi\rangle$ can be replaced in all quantum mechanical calculations,
without any loss of generality,  by the corresponding dyadic operators
$|\psi\rangle\langle\varphi_0|$, for any {\it fixed} vector $|\varphi_0\rangle$ of unit length,
$\langle\varphi_0|\varphi_0\rangle =1$.
For example, we can superpose dyadic operators,
\be
\alpha |\psi_1\rangle +\beta |\psi_2\rangle \longleftrightarrow
\alpha |\psi_1\rangle\langle\varphi_0|  +\beta |\psi_2\rangle \langle \varphi_0|\,;
\label{superpose}
\ee
we can evolve them in time using Schr\"odinger's equation,
\be
i \partial_t |\psi(t)\rangle={\hat H}|\psi(t)\rangle
\longleftrightarrow
i\partial_t |\psi(t)\rangle\langle\varphi_0|={\hat H}
|\psi(t)\rangle\langle  \varphi_0|\,;
\label{schro1}
\ee
we can use them to calculate corresponding pure-state density operators,

\bea
\R=|\psi\rangle\langle\psi|= |\psi\rangle\langle\varphi_0|(|\psi\rangle\langle\varphi_0|)^{\dagger}\,;
\label{rho_calc}
\eea
we can use them to calculate expectation values,
\be
\langle {\hat A}\rangle = \langle \psi |{\hat A}|\psi\rangle =
{\rm Tr}\left(\left(|\psi\rangle\langle\varphi_0|\right)^\dagger {\hat A}
|\psi\rangle\langle\varphi_0|\right)\,;
\label{expec1}
\ee
and, most important, we can use them to calculate transition amplitudes, 
\be
\langle\psi_1|\psi_2\rangle=
{\rm Tr} \left(\left(|\psi_1\rangle\langle \varphi_0|\right)^\dagger | \psi_2\rangle
\langle\varphi_0 | \right)\,,
\label{transition1}
\ee
and not just transition {\it probabilities} $|\langle\psi_1|\psi_2\rangle|^2$.  
Transition probabilities, but not transition amplitudes, can
readily be determined from
the density matrix or, equivalently,  from the Wigner function.

\vs\ni
It is now a simple matter \cite{smith1} to combine  two ideas --- the notion of replacing state vectors $|\psi\rangle$ 
by dyadic operators $|\psi\rangle\langle \varphi_0 |$, and the notion of  mapping operators into functions on phase space 
using the Weyl-Wigner transform  --- in order to define phase space amplitudes $\Psi(q,p)$.  We simply set

\bea
\Psi(q,p) =\frac{1}{\sqrt{2\pi}}\W \left(\,|\psi\rangle\langle\varphi_0|\,\right)(q,p)
\equiv \frac{1}{\sqrt{2\pi}}{\rm Tr }\left(|\psi\rangle\langle \varphi_0|\Del (q,p)\right) \,.
\label{Psi_def}
\eea
Here $\Del (q,p)$ is the Weyl-Wigner-Stratonovich kernel operator 
\cite{weyl,wigner,stratonovich,grossmann,royer,brif,lobo,ellinas}, given 
in terms of $q$, $p$ 
and corresponding (dimensionless) canonical operators $\qO$, $\pO$, by   

\bea
\Del (q,p)=2\,e^{2i(p\,\,\qO -q\,\pO)}\,\Par\,,
\label{Del_def}
\eea
where $\Par$ is the parity operator, satisfying

\bea
\qO\,\Par=-\Par\,\qO\,,\quad \pO\,\Par=-\Par\,\pO\,,\quad \Par^2=1.  
\label{parity}
\eea  
\vs\ni
We recall at this point that the star product \cite{neumann,groenewold,moyal} of phase space functions is 
defined through the Weyl-Wigner transform by

\bea
&\,& \W(\A)=A\,,\,\, \W(\B)=B\,\,\Rightarrow\,\,\W(\A\B)=A\star B\,,
\label{star1}
\eea
leading to
\bea
(A\star B)(q,p)
&=&A(q,p)\,e^{iJ/2}\,B(q,p)\,,\quad (J=\leftder_q\rightder_p-\leftder_p\rightder_q )
\mea\mea
&=& e^{i(\partial_q\partial_{p'}-\partial_p\partial_{q'})/2}\,A(q,p)B(q',p')|_{(q',p')=(q,p)}\,.
\label{star2}
\eea
The expressions involving differential operators are well-defined if
$A$ and $B$ are polynomial functions.  In more general cases, they define asymptotic expansions,

\bea
A(q,p)\star B(q,p)=A(q,p)B(q,p)+\half iA(q,p)\,J\,B(q,p)+{\rm O}(2)\,,
\label{asymptotic}
\eea
provided $A$ and $B$ represent observables that are asymptotically regular at $\hbar =0$ \cite{osborn}.  
In (\ref{asymptotic}),  O(2) denotes terms of second order in Planck's constant (here implicit).

\vs\ni 
A string of results now follows from (\ref{Psi_def}) and (\ref{star1}-\ref{asymptotic}):

\vs\ni
\begin{enumerate}
 \item {\em Complex phase space amplitudes}: For each choice of window vector $|\varphi_0\rangle$, 
a (distinct) set of amplitudes $\Psi$ 
is defined by (\ref{Psi_def}), corresponding to the set of all state vectors $|\psi\rangle$.  
Each $\Psi$ therefore carries implicitly  a label $\varphi_0$, 
which we suppress.  
For any choice of $|\varphi_0\rangle$, the amplitudes $\Psi$ are complex-valued in general, and carry the same phases as the 
corresponding state vectors $|\psi\rangle$, up to a constant phase shift determined by $|\varphi_0\rangle$.

\vs\ni \item
{\em Expectation values}: Corresponding to (\ref{expec1}), expectation values can be expressed in the form

\bea
\langle{\widehat A}\rangle &=&\int \overline{\Psi(q,p)}\star A(q,p)\star \Psi(q,p)\,d\Gamma
\mea\mea
&=&\int \left\{\Psi(q,p)\star \overline{\Psi(q,p)}\right\} A(q,p)\,d\Gamma
\mea\mea
&=&  \int \overline{\Psi(q,p)}\left\{A(q,p)\star \Psi(q,p)\right\}\,d\Gamma\,.
\label{expec2}
\eea
Here we have used the result \cite{kubo,fairlie}

\bea
{\rm Tr}(\A\B)\longrightarrow \int A(q,p)\star B(q,p)\,d \Gamma =\int A(q,p)B(q,p)\,d\Gamma\,,
\label{trace_formula}
\eea 
for suitably smooth functions $A$ and $B$. 
\vs\ni
\item
 {\em Transition probabilities}: Corresponding to (\ref{transition1}), transition amplitudes and subsequently transition probabilities
can be calculated from

\bea
\langle\psi_1|\psi_2\rangle=\int \overline{\Psi_1(q,p)}\Psi_2(q,p)\,d\Gamma\,.
\label{transition3}
\eea
In particular,

\bea
\langle\psi|\psi\rangle=\int \overline{\Psi(q,p)}\Psi(q,p)\,d\Gamma=1\,
\label{Psi_normalization1}
\eea
which suggests that $|\Psi(q,p)|^2$, like $W(q,p)$, can be regarded as a quasi--  
probability distribution over $\Gamma$ in its own right.  We shall see below that in fact
$|\Psi(q/2,p/2)/2|^2$ is a generalized Husimi distribution.

\vs\ni
\item {\em Generalized Born interpretation}: Corresponding to (\ref{rho_calc}), we have

\bea
W(q,p)=\Psi(q,p)\star \overline{\Psi(q,p)}\,,
\label{pseudoBorn}
\eea
which, bearing (\ref{expec2}) in mind, we refer to as the Born interpretation of $\Psi$, 
by analogy with the relation between the
wavefunction and probability density in configuration space. 
Then (\ref{asymptotic}) suggests the expansion

\bea
W(q,p)=\Psi(q,p)\overline{\Psi(q,p)}+\half i \Psi(q,p)\,J\,\overline{\Psi(q,p)}+ \dots\,,
\label{approx1}
\eea
which in turn suggests a non-negative approximation to the Wigner function,

\bea
W(q,p)\approx |\Psi(q,p)|^2\geq 0\,.
\label{approx2}
\eea
However, (\ref{approx1}) and (\ref{approx2}) must be treated with caution because $\Psi$ and $W$ may 
not be asymptotically regular  at $\hbar=0$ \cite{osborn,berry}. 
Comparisons of $|\Psi|^2$ with $W$ in examples that follow illustrate the difficulty  --- see for example
(\ref{cs12}) and (\ref{cs13}).

\vs\ni
 \item {\em Superposition property}: For each choice of $|\varphi_0\rangle$, the mapping from $|\psi\rangle$ to $\Psi$ is linear and phase space amplitudes can be superposed,

\bea
|\psi_{12}\rangle=c_1|\psi_1\rangle+c_2|\psi_2\rangle \longleftrightarrow \Psi_{12}(q,p)=c_1\Psi_1(q,p)+c_2\Psi_2(q,p)\,,
\label{linearity}
\eea
preserving the phase relations between state vectors.

\vs\ni
Note from (\ref{pseudoBorn}) that if $W_1$, $W_2$ and $W_{12}$ are the Wigner functions 
corresponding to the 
states and phase space amplitudes in (\ref{linearity}),  then

\bea
W_{12}= \left(c_1\Psi_1 +c_2\Psi_2\right)\star \overline{\left(c_1\Psi_1 +c_2\Psi_2\right)}
\mea\mea
= |c_1|^2 W_1 +|c_2|^2 W_2 \qquad\qquad\qquad\qquad
\mea\mea
+c_1\overline{c_2}\,\Psi_1\star\overline{\Psi_2}+c_2\overline{c_1}\,\Psi_2\star\overline{\Psi_1}\,.
\label{wigner_superpose}
\eea
In contrast, it is possible \cite{bracken5} but not straightforward to express $W_{12}$, for example,  
directly in terms of $W_1$ and $W_2$.
\vs\ni
Similarly for the generalized Husimi distribution, we have

\bea
|\Psi_{12}|^2= \left(c_1\Psi_1 +c_2\Psi_2\right)\, \overline{\left(c_1\Psi_1 +c_2\Psi_2\right)}
\mea\mea
= |c_1|^2 |\Psi_1|^2 +|c_2|^2 |\Psi_2|^2 \qquad\qquad\qquad\qquad
\mea\mea
+c_1\overline{c_2}\,\Psi_1\overline{\Psi_2}+c_2\overline{c_1}\,\Psi_2\overline{\Psi_1}\,.
\label{husimi_superpose}
\eea

\vs\ni
 \item {\em Normalization and subspace of amplitudes}: The results (\ref{transition3}) 
and (\ref{Psi_normalization1}) show that, for whatever choice of $|\varphi_0\rangle$, 
the complex amplitude $\Psi$ belongs to the Hilbert space  of  square-integrable 
functions on phase space, and is normalized when $|\psi\rangle$ is normalized.  
In fact all such amplitudes for a given window state 
lie in a proper closed subspace $\subspace$ of that Hilbert space, 
characterized by the property that

\bea
\Psi(q,p)\star W_{\varphi_0}(q,p)=\Psi(q,p)/2\pi\,,
\label{Psi_charac1}
\eea
where $W_{\varphi_0}$ is the Wigner function corresponding to the state $|\varphi_0\rangle$.  This
follows from the identity

\bea
\psistate\langle\varphi_0|\varphi_0\rangle\langle\varphi_0|=\psistate\langle\varphi_0|
\label{Psi_charac2}
\eea
under the action of $\W$.
Note also the identity

\bea
W_{\psi}(q,p)\star\Psi(q,p)=\Psi(q,p)/2\pi\,,
\label{Psi_charac3}
\eea
which follows from

\bea
|\psi\rangle\langle\psi|\psi\rangle\langle\varphi_0|=|\psi\rangle\langle\varphi_0|\,.
\label{Psi_charac4}
\eea
\vs\ni
It can be seen from (\ref{transition3}) that if $\{|\psi_n\rangle\}$ is a complete 
orthonormal set of vectors in $\Hil$, then the corresponding
set of phase space amplitudes $\{\Psi_n\}$ provides a complete orthonormal set in 
$\subspace$ under the scalar product given by the RHS of (\ref{transition3}).  
Thus $\subspace$ is a Hilbert space in its own right, which for each choice of window 
provides an image of $\Hil$ within the space of 
square-integrable functions on $\Gamma$.  It is  clear \cite{ascensi} that $\Hil$, 
which after all is commonly realized in the coordinate representation 
as the space of square-integrable functions on the real {\em line}, 
has uncountably many images in
the space of square-integrable functions on the phase {\em plane}.  
 
\end{enumerate}

\section{A class of integral transforms}
In the coordinate representation, formula (\ref{Psi_def}) takes the form 

\bea
\Psi(q,p)&=&\frac{1}{\sqrt{2\pi}}\int \psi (q-\half y)\overline{\varphi_0(q+\half y)}\, e^{ipy} \,dy
\mea\mea
&=&\sqrt{2/\pi}\int \psi (u)\overline{\varphi_0(2q-u)}\, e^{2ip(q-u)} \,du
\mea\mea
&=&\sqrt{2/\pi}\int \psi (2q-u)\overline{\varphi_0(u)}\, e^{2ip(u-q)} \,du\,,
\label{Psi_def2}
\eea
where $\psi$ and $\varphi_0$ are the wavefunctions corresponding to $\psistate$ and $|\varphi_0\rangle$ respectively.  
\vs\ni
The first of these formulas shows that, 
apart from a normalization factor, $\Psi$ is the `two-state' or `two-sided' Wigner function \cite{zachos,hillery} that is sometimes denoted
$W_{\psi\varphi_0}(q,p)$.  But as the notation $\Psi$ suggests,
we consider it now as varying with, and determined by 
$|\psi\rangle$, with $|\varphi_0\rangle$ held fixed once and for all. We emphasize in particular 
that when $|\psi(t)\rangle$ evolves in time,  $\Psi(t)$ evolves also 
with {\em only the LH member} of $|\psi(t)\rangle\langle\varphi_0|$ changing (see below). 

\vs\ni
The second formula (\ref{Psi_def2}) shows that the  transform from $\psi$ to $\Psi$, obtained with any fixed choice 
of the normalized wavefunction $\varphi_0$, is closely related to Gabor's `windowed Fourier transform' \cite{gabor,cohen1,cohen2}, 
which is

\bea
\Phi(q,p)=\frac{1}{\sqrt{2\pi}}\int \psi(u) w(u-q)\,e^{-ipu}\,du\,,
\label{window_trasform}
\eea
with $w(x)$ the window function.  Choosing $w(x)=\overline{\varphi_0(-x)}$ then leads to

\bea
\Psi(q/2,p/2)=2\,e^{iqp/2}\,\Phi(q,p)\,.
\label{transform_relation}
\eea
It is easily checked that $\Psi$ and $e^{iqp/2}\,\Phi$ are also related by 
a `symplectic Fourier transform',

\bea
e^{i qp/2}\,\Phi(q,p)=\frac{1}{2\pi}\,\int \Psi(q',p')\,e^{i(p'q-q'p)}\,d\Gamma'\,.
\label{transform_relation1}
\eea
Gabor  initially chose a simple Gaussian centred on the origin for the window function $\varphi_0(x)$.  
In the present context, this choice leads to Bargmann's representation \cite{bargmann} of the wavefunction as an entire function on $\Gamma$, 
regarded as the complex plane (see below).  

\vs\ni
Note also that the second formula (\ref{Psi_def2}) shows each transform in the general form

\bea
\Psi(q,p)=\int T_{\varphi_0}(q,p,u)\psi(u)\,du\,,
\mea\mea
 T_{\varphi_0}(q,p,u)= \sqrt{2/\pi}\,\overline{\varphi_0(2q-u)}e^{2ip(q-u)}\,.
\label{transform1}
\eea

\vs\ni
From (\ref{Psi_def}) we can also write

\bea
\Psi (q,p) ={\rm Tr} (\psistate\langle\varphi_0|\Del(q,p))&=&\langle\varphi_0|\Del (q,p)|\psi\rangle =\langle {\mathcal T}_{\varphi_0}(q,p)\psistate\,,
\mea\mea
|{\mathcal T}_{\varphi_0}(q,p)\rangle&=&\Del(q,p)|\varphi_0\rangle\,. 
\label{transform2}
\eea
\vs\ni
In a similar way it can be seen that

\bea
\Phi(q,p)&=&\langle {\mathcal Q}_{\varphi_0}(q,p)\psistate\,,
\mea\mea
|{\mathcal Q}_{\varphi_0}(q,p)\rangle&=&e^{i(p\,\,\qO -q\,\pO)}\,|\varphi_0\rangle\,. 
\label{transform_relation2}
\eea
The transform inverse to (\ref{Psi_def2}) is obtained by considering $\W^{-1}\left(\Psi\right)$
in the coordinate representation, which leads to

\bea
\frac{1}{2\pi}\int\Psi((x+y)/2, p)\,e^{ip(x-y)}\,dp
&=&\langle x|\psi\rangle\langle\varphi_0|y\rangle \,,
\mea\mea
&=& \psi(x) \overline{\varphi_0(y)}\,.
\label{inverse_transform1}
\eea
Hence

\bea
\psi(x)=\frac{1}{2\pi \overline{\varphi_0(y)}}\int\Psi((x+y)/2, p)\,e^{ip(x-y)}\,dp
\label{inverse_transform2}
\eea
wherever $\varphi_0(y)\neq 0$, and also

\bea
\psi(x)= \frac{1}{2\pi}\int\!\!\!\int \Psi((x+y)/2, p)\varphi_0(y)\,e^{ip(x-y)}\,dy\,dp\,.
\label{inverse_transform3}
\eea
Using the definition of $\W^{-1}$ in terms of $\Del(q,p)$ \cite{ellinas}, 
the result (\ref{inverse_transform3}) can also be written as

\bea
|\psi\rangle=\frac{1}{2\pi}\int \Psi(q,p)|{\mathcal T}_{\varphi_0}(q,p)\rangle\,d\Gamma\,.
\label{inverse_transform4}
\eea
Similarly

\bea
|\psi\rangle=\frac{1}{2\pi}\int \Phi(q,p)|{\mathcal Q}_{\varphi_0}(q,p)\rangle\,d\Gamma\,.
\label{inverse_transform5}
\eea

\section{Generalized Husimi functions}

\vs\ni
The first of formulas (\ref{Psi_def2}) says that $\Psi(q,p)$ is the Fourier transform of 

\bea
A_q(y)=\psi(q-\half y)\,\overline{\varphi_0(q+\half y)}\,.
\label{husimi1}
\eea
Similarly, $\overline{\Psi(q,p)}$ is the Fourier transform of 

\bea
B_q(y)=\overline{A_q(-y)}\,,
\label{husimi2}
\eea
and so $|\Psi(q,p)|^2$ is the product of these transforms. By the convolution theorem for Fourier transforms \cite{koerner}, it follows that

\bea
|\Psi(q,p)|^2
=\frac{1}{2\pi}\int A_q(y-z)B_q(z)\,e^{ipy}\,dz\,dy&\,&
\mea\mea
=\frac{1}{2\pi}\int \psi(q-\half y+\half z)\overline{\psi(q+\half z)}\qquad\qquad\qquad&\,& 
\mea\mea
\times \,\overline{\varphi_0(q+\half y -\half z)}
\varphi_0(q-\half z)\,e^{ipy}\,dz\,dy
\mea\mea
=\frac{2}{\pi}\int\psi(u+\half v)\overline{\psi(u-\half v)}\qquad\qquad\qquad&\,&
\mea\mea
\times\,\overline{\varphi_0(2q-u-\half v)}\varphi_0(2q-u+\half v)\,e^{-2ipv}\,du\,dv\,.
\label{husimi3}
\eea
Setting

\bea
f(r,v)=\int\overline{\varphi_0(\theta-\half v)}\,\varphi_0(\theta +\half v)\,e^{ir\theta }\,d\theta\,,
\label{husimi4}
\eea
so that

\bea
\overline{\varphi_0(\theta-\half v)}\,\varphi_0(\theta +\half v)=
\frac{1}{2\pi}\int f(r,v)\,e^{-ir\theta}\,dr\,,
\label{husimi5}
\eea 
we then have

\bea
|\Psi(q/2,p/2)|^2=\frac{1}{\pi^2}\int \psi(u+\half v)\overline{\psi(u-\half v)} 
f(r,v)e^{-ir(q-u)}\,e^{-ipv}\,dr\,du\,dv\,,
\label{husimi6}
\eea
showing that $|\Psi(q/2, p/2)/2|^2$ is a (nonnegative) distribution function 
from Cohen's general class \cite{cohen1,claasen}. Note from (\ref{transform_relation}) that

\bea
|\Psi(q/2, p/2)/2|^2 \, d(q/2)\, d(p/2)=|\Phi(q,p)|^2\,dq\,dp\,.
\label{husimi6a}
\eea
In the signals literature,  $|\Phi(q,p)|^2$ is known as the spectrogram of the signal $\psi$. 
It can be regarded as a generalized Husimi function \cite{smith2} (see below).  
In particular, when the window function is the simple Gaussian
corresponding to the oscillator ground state (see below), 

\bea
\varphi_0(x)=e^{-x^2/2}/\pi^{1/4}\,,
\label{husimi7}
\eea
we get from (\ref{husimi4})
\bea
f(r,v)=e^{-(r^2+v^2)/4}\,,
\label{husimi8}
\eea
so that 

\bea
|\Psi(q/2, p/2)|^2\, d(q/2)\, d(p/2 \left[=|\Phi(q,p)|^2\,dq\,dp\right]=Q(q,p)\,dq\,dp\,,
\label{husimi9}
\eea
where $Q$ is the original Husimi function \cite{husimi}.
\vs\ni
It is remarkable, especially in view of (\ref{approx1}) and (\ref{approx2}),  
that $\Psi(q,p)\star \overline{\Psi(q,p)}$ equals the Wigner function for the state $|\psi\rangle$,
while $\Psi(q/2, p/2)\,\overline{\Psi(q/2, p/2)}$ defines a generalized Husimi function for that same state, whatever the choice of window.  
In this way $\Psi\star \overline{\Psi}$ and $\Psi\, \overline{\Psi}$ define different aspects of the generalized Born interpretation of $\Psi$ (see above).

\section{Gaussian windows}
 
 \vs\ni
Amongst the simplest choices for a window state is a squeezed state,  described by a 
Gaussian function in the coordinate representation,

\bea
\varphi_0(x)=(\beta^2/\pi)^{1/4}\,e^{-\beta^2(x-x_W)^2/2+ik_W(x-x_W/2)}
\mea\mea
=(\beta^2/\pi)^{1/4}\,e^{\beta^2\lambda(\lambda-{\bar \lambda})/4}\,e^{-\beta^2(x-\lambda)^2/2}\,,\quad \lambda=x_W+ik_W/\beta^2\,,
\label{gaussian5}
\eea
where $\beta$, $x_W$ and $k_W$ are  real constants.  This has 
momentum space representation (Fourier transform)

\bea
\tilde{\varphi}_0(k)=(1/\beta^2\pi)^{1/4}\,e^{-(k-k_W)^2/2\beta^2-i(k-k_W/2)x_W}
\mea\mea
= (1/\beta^2\pi)^{1/4}\,e^{-\beta^2\lambda(\lambda+{\bar \lambda})/4}\,e^{-(k+i\beta^2\lambda)^2/2\beta^2}\,.
\label{gaussian6}
\eea 
 Then

\bea
|\varphi_0(x)|^2= e^{-\beta^2(x-x_W)^2}\,,\quad |\tilde{\varphi_0}(k)|^2= e^{-(k-k_W)^2/\beta^2}\,,
\label{gaussian7}
\eea
which are peaked at $x=x_W$ and $k=k_W$ respectively.  
Note also that $\varphi_0(x)$ is the coordinate space representation of a solution of

\bea
\frac{1}{\sqrt{2}}(\beta  \qO+i\pO/\beta)|\varphi_0\rangle=\frac{\beta\lambda}{\sqrt{2}} |\varphi_0 \rangle\,.
\label{cs1}
\eea
Before proceding, we recall that 
\bea
\W(\qO^n)=q\star q\star q\dots \star q =q^n\,,\quad \W(\pO^n)=p^n\,,
\label{qp_powers}
\eea 
and that, following Bopp \cite{bopp,kubo}

\bea
\W\left(\A\,\qO^n\right)(q,p)=A(q,p)\star q^n=[q-\half i\partial_p]^n A(q,p)\equiv[\qBR]^n A(q,p)\,,\mea\mea
\W\left(\A\,\pO^n\right)(q,p)=A(q,p)\star p^n= [p+\half i\partial_q]^n A(q,p)\equiv[\pBR]^n A(q,p)\,,
\mea\mea
\W\left(\qO^n\,\A\right)(q,p)=q^n\star A(q,p)= [q+\half i\partial_p]^n A(q,p)\equiv[\qBL ]^n A(q,p)\,,
\mea\mea
\W\left(\pO^n\,\A\right)(q,p)=p^n\star A(q,p)= [p-\half i\partial_q]^n A(q,p)\equiv[\pBL]^n A(q,p)\,.
\label{star3}
\eea
Applying the first two of these results to $\A=|\psi\rangle\langle\varphi_0|$ with $n=1$, 
it follows from (\ref{cs1}) that every phase space amplitude $\Psi$ constructed with the Gaussian window state 
(\ref{gaussian5}) must satisfy

\bea
\frac{1}{\sqrt{2}}(\beta\qBR -i\pBR/\beta)\Psi(q,p)=\frac{\beta\bar{\lambda}}{\sqrt{2}}\Psi(q,p)\,,
\label{cs2}
\eea
and it is easily seen that the general solution of this equation has the form

\bea
\Psi(q,p)=\sqrt{2/\pi}\,e^{-\bar{z}z/2+\beta(z+{\bar z})\bar{\lambda}/\sqrt{2}}\,G_{\psi}(z)\,,\quad\!\!\!
z=\sqrt{2}(\beta q-ip/\beta)\,,
\label{cs3}
\eea
where $G_{\psi}$ is arbitrary and   we have included an extra  $z$-dependent prefactor to $G_{\psi}$ for later convenience.  
The form of  $G_{\psi}(z)$ is determined by $\psi$ from (\ref{gaussian5}) and the second formula in (\ref{Psi_def2}), leading to

\bea
G_{\psi}(z)=(\beta^2/\pi)^{1/4}\,e^{\beta^2\bar{\lambda}(\bar{\lambda}-\lambda)/4}\qquad\qquad\qquad
\mea\mea
\times \int e^{-(z^2+\beta^2(u+\bar{\lambda})^2-2\sqrt{2}\beta z u)/2}\,\psi(u)\,du\,.
\label{G_transform1}
\eea

\section{Generalized Bargmann representation}
\vs\ni
Because the exponential factor in (\ref{cs3}) is independent of $\psi$, this result establishes a 
generalized Bargmann representation \cite{bargmann},

\bea
|\psi\rangle \longleftrightarrow G_{\psi}(z)\,,
\label{gen_bargmann1}
\eea
with $G_{\psi}$ an analytic function of the complex variable $z$ (no $\bar{z}$-dependence),
or equivalently of $p+i\beta^2 q$, 
in a Hilbert space $\Hil_G$ with weighted scalar product, from (\ref{transition3}) and (\ref{cs3}), 

\bea
\left(G_{\psi_1}\,,\,G_{\psi_2}\right)=
(2/\pi)\,\int \overline{G_{\psi_1}(z)}\,G_{\psi_2}(z)\,e^{-\bar{z}z+\beta (\lambda+\bar{\lambda})(z+\bar{z})/\sqrt{2}}\,d^2z\,.
\label{gen_bargmann2}
\eea
Here $d^2z=dq\,dp$.
The creation and  annihilation  operators  $(\qO \mp i\pO)/\sqrt{2}$ acting on each $|\psi\rangle$ in $\Hil$, are represented on $\subspace$ as

\bea
(\qBL\mp i\pBL)/\sqrt{2} \,\equiv \,\pm \half(\beta\pm 1/\beta)(\half z-\partial_{\bar{z}})\mp\half(\beta\mp 1/\beta)(\half \bar{z}+\partial_z)\,,
\nonumber\\
\label{cs4}
\eea
 and therefore, taking into account the exponential factor in (\ref{cs3}), they are represented on $\Hil_G$ as    

\bea
(\qO-i\pO)/\sqrt{2}\longleftrightarrow \sigma z -\tau\partial_z-\beta^2\bar{\lambda}/\sqrt{2}
\mea\mea
(\qO+i\pO)/\sqrt{2}\longleftrightarrow \sigma\partial_z-\tau z+\beta^2\bar{\lambda}/\sqrt{2}\,,
\mea\mea
\sigma=(\beta^2+1)/2\beta\,,\quad \tau = (\beta^2-1)/2\beta\,.
\label{gen_bargmann3}
\eea 
Then

\bea
\qO&\longleftrightarrow &(z+\partial_z)/\beta\sqrt{2}
\mea
\mea
\pO&\longleftrightarrow &i\beta(z-\partial_z)/\sqrt{2} - i \beta^2{\bar \lambda}\,. 
\label{gen_bargmann4}
\eea
Note that the RHSs of (\ref{gen_bargmann3}) and (\ref{gen_bargmann4}) 
do not involve $\bar{z}$ and so preserve the analyticity of 
any $G_{\psi}$ upon which 
they act.  They satisfy the canonical commutation relations and have the appropriate 
hermiticity properties with respect to the scalar product (\ref{gen_bargmann2}).  
These expressions generalize those of the usual 
Bargmann representation, which corresponds to the case $\beta=1$ and $\lambda=0$, 
that is, the case of a Gaussian window 
(\ref{gaussian5}) with the same scaling as that used for $\qO$ and $\pO$, and centered on $x=0$.
Then $\sigma =1$ and $\tau =0$, and the expressions on the RHS of 
(\ref{gen_bargmann3}) reduce to the familiar $z$ and $\partial_z$, respectively.   
Furthermore, it can be seen that (\ref{G_transform1})  
reduces when $\beta=1$ and $\lambda =0$ to Definition (2.3) of the Bargmann transform in \cite{bargmann}.
\vs\ni
The intimate connection between the usual Bargmann wavefunction 
and the Wigner function, and with the phase space formulation of quantum mechanics 
more generally, has been discussed previously \cite{janssen,wunsche,parisio} from various 
other points of view.  Furthermore, 
it has recently been shown \cite{ascensi} in the context of the Gabor transform, 
that only Gaussian windows give rise to spaces of  analytic functions of the (generalized) Bargmann type.

\section{A test state and its phase space amplitude}
\vs\ni
Consider  a  normalized test wavefunction 

\bea
&\,&\psi (x)= \frac{1}{\sqrt{\sqrt{\pi}(1+2\sqrt{2})}} \,\left( e^{-(x-1)^2/2}+4ix \,e^{-x^2}\right)
\mea
\longleftrightarrow&\,&
\mea
&\,&\tilde{\psi}(k)=\frac{1}{\sqrt{\sqrt{\pi}(1+2\sqrt{2})}} \,\left(e^{-k^2/2-ik}+\sqrt{2}k\,e^{-k^2/4}\right)\,.
\label{test1}
\eea
which has as Wigner function

\bea
W(q,p)= \frac{1}{\pi(1+2\sqrt{2})}\,\left\{\frac{}{}e^{-(q-1)^2-p^2}\qquad\qquad\qquad\qquad\right.
\mea\mea
\left. +2\sqrt{2}(4q^2+p^2-1)\,e^{-2q^2-p^2/2}
-\frac{8\sqrt{2}}{3\sqrt{3}}\,\left[(2q-1)\sin(2p\,[q+1]/3)\right.\right.
\mea\mea
\left. \left.-2p\,\cos(2p\,[q+1]/3)\right.]\,e^{-(4q^2-4q+2p^2+1)/3}\,\,\right\}
\label{test2}
\eea
and which, for the choice (\ref{gaussian5}) as window with $x_W=\langle \qO\rangle$ and $k_W=\langle \pO\rangle$, leads to the phase space amplitude

\bea
\Psi(q,p)=N\,\left\{e^{-[4\beta^2q^2+4p^2+4i(\beta^2-1)qp-4\beta^2(\bar{\lambda}+1)q]/2(\beta^2+1)}\right.
\mea\mea
\times e^{[4i(1-\beta^2\bar{\lambda})p+\beta^2(\bar{\lambda}+1)^2]/2(\beta^2+1)}/(\beta^2+1)^{1/2}
\mea\mea
+4i(2\beta^2q-2ip-\beta^2\bar{\lambda})\,e^{-[4\beta^2q^2+2p^2+2i(\beta^2-2)qp]/(\beta^2+2)}
\mea\mea
\left.\times e^{[4\beta^2\bar{\lambda}q+2i\beta^2\bar{\lambda}p-\beta^2\bar{\lambda}^2]/(\beta^2+2)}/(\beta^2+2)^{3/2}\,\,\right\}\,.
\label{test3}
\eea
Here 

\bea
N=\sqrt{4\beta/\pi(1+2\sqrt{2})}\,\,e^{\beta^2\bar{\lambda}(\bar{\lambda}-\lambda)/4}\,,\qquad \lambda=x_W+ik_W/\beta^2\,,
\mea\mea
x_W=\langle \qO\rangle = 1/(1+2\sqrt{2}) \,,\qquad k_W=\langle \pO\rangle =8\sqrt{2}/9\,e^{1/3}\sqrt{3}(1+2\sqrt{2})\,.
\label{window_means1}
\eea
The values of $x_W$ and $k_W$ in the definition (\ref{gaussian5}) of
the window function have been equated to the expectation values $\langle \qO\rangle$, $\langle\pO\rangle$ of position and 
momentum for the state (\ref{test1}), 
so that $|\varphi_0(x)|$ and $|\tilde{\varphi}_0(k)|$ are  localized near these values in position and momentum space, respectively, in accordance with (\ref{gaussian7}). 

\vs\ni
Fig. 1 shows the real and imaginary parts of $\Psi$ as in (\ref{test3}) when $\beta=1$.  Fig. 2 shows the Wigner function $W$ of (\ref{test2}), 
which is not everywhere positive, and 
the 
distribution $|\Psi|^2$  in this case, for $\Psi$ as in (\ref{test3}), 
with $\beta =1$.  Fig. 3 shows $|\Psi|^2$ for $\beta=0.5$  and $2$.  Comparison of the subplots shows the influence of  the uncertainty principle: 
when $\beta^2\ll 1$, the structure of $|\Psi |^2$  better delineates the $p$-dependence of $W$, 
while when $\beta^2 \gg 1$, it better delineates the $q$-dependence.  

\vs\ni
[Figs. 1, 2 and 3 near here.]
\vs\ni

\section{Coherent states in phase space}

\vs\ni
The formulas (\ref{transform2}) and (\ref{transform_relation2}) show that 
$\Psi$ and $\Phi$ can always be regarded as
generalized coherent states in Perelemov's sense \cite{perelemov}.  However,
we now ask what phase space amplitudes $\Psi$ correspond to the familiar 
coherent state in $\Hil$  defined up to a constant phase by

\bea
\frac{1}{\sqrt{2}}(\qO+i\pO)|\psi_{\mu}\rangle=\frac{\mu}{\sqrt{2}}|\psi_{\mu}\rangle\,,
\quad \mu=x_C+ik_C\in {\mathbb C}
\label{cs5}
\eea
and normalization.  Here $x_C$ and $k_C$ are arbitrary real numbers.  
It follows from (\ref{star3}) that each such state has a corresponding 
phase space amplitude $\Psi_{\mu}$ satisfying

\bea
\frac{1}{\sqrt{2}}(\qBL+i\pBL)\Psi_{\mu}(q,p)&\equiv &\frac{1}{\sqrt{2}}(q+\half i\partial_p+ip+\half\partial_q)\Psi_{\mu}(q,p)
\mea\mea
&=&\frac{\mu}{\sqrt{2}}\Psi_{\mu}(q,p)\,,
\label{cs6}
\eea
and hence having the general form

\bea
\Psi_{\mu}(q,p)=K_{\mu}({\bar w})\,e^{-{\bar w}w/2}\,e^{\mu w/\sqrt{2}}\,,\quad w=\sqrt{2}(q-ip)\,.
\label{cs7}
\eea
Here the precise form of $K_{\mu}$ is determined by the phase and normalization of $|\psi_{\mu}\rangle$ 
and the choice of window $|\varphi_0\rangle$.  
Each choice of $|\varphi_0\rangle$
leads also to a  corresponding subspace $\subspace$ of square integrable functions on $\Gamma$ through 
(\ref{Psi_charac1}) and (\ref{Psi_charac2}), 
and it is remarkable that the phase space amplitudes (\ref{cs7}) --- 
which we may call coherent states in phase space --- must, for every choice of window state,  
form an overcomplete set in the corresponding $\subspace$ for varying $\mu$, 
just as the coherent states  $|\psi_{\mu}\rangle$ form an overcomplete set in $\Hil$.  
\vs\ni
If the window  is a Gaussian as in (\ref{gaussian5}), then the amplitude $\Psi_{\mu}$ must also satisfy (\ref{cs2}), 
leading from (\ref{cs7}) to

\bea
K_{\mu}({\bar w})={\rm const.}\,e^{-\tau {\bar w}^2/2\sigma}\,e^{(\tau\mu +\beta{\bar \lambda}){\bar w}/\sqrt{2}\sigma}\,,
\label{cs8}
\eea
where $\sigma$ and $\tau$ are as in (\ref{gen_bargmann3}).    
In this case we know also that $\Psi_{\mu}$ must be of the form (\ref{cs3}), and we find

\bea
\Psi_{\mu}(q,p)=V_{\mu}(z)\,e^{-{\bar z}z/2}\,e^{\beta {\bar \lambda}{\bar z}/\sqrt{2}}\,,\quad z=\sqrt{2}(\beta q-ip/\beta)\,,
\mea\mea
V_{\mu}(z)={\rm const.}\,e^{\tau z^2/2\sigma}\,e^{(\mu-\tau\beta{\bar \lambda})z/\sigma\sqrt{2}}\,.\qquad\qquad\qquad
\label{cs9}
\eea
Consistency of (\ref{cs7}), (\ref{cs8}) and (\ref{cs9}) is easily checked, leading to

\bea
\Psi_{\mu}(q,p)={\rm const.} \,e^{-2[\beta^2q^2+p^2+i(\beta^2-1)qp-\beta^2
({\bar \lambda}+\mu)q-i(\beta^2{\bar\lambda}-\mu)p]/(\beta^2+1)}\,,
\label{cs10}
\eea
and hence

\bea
\overline{\Psi_{\mu}(q,p)} \,\Psi_{\mu}(q,p)=\frac{4\beta}{\pi(\beta^2+1)}\,e^{-4[\beta^2(q-\eta)^2+(p-\zeta)^2]/(\beta^2+1)}\,,
\mea\mea
\eta=(x_W+x_C)/2\,,
\quad
\zeta=(k_W+k_C)/2\,,
\label{cs11}
\eea
with $x_W$, $k_W$ as in (\ref{gaussian5}) and $x_C$, $k_C$ as in (\ref{cs5}).
In (\ref{cs11}), we have taken the normalization condition (\ref{Psi_normalization1}) into account.  
On the other hand, the Wigner function corresponding to the coherent state $|\psi_{\mu}\rangle$ is \cite{hillery}

\bea
W_{\mu}(q,p)=\frac{1}{\pi}\,e^{-(q-x_C)^2-(p-k_C)^2}\,.
\label{cs12}
\eea
If we choose the Gaussian window to be centered on the same coordinate and momentum values $x_C$, $k_C$
as the Wigner function, with the same choice of length scale --- that 
is to say, if we choose $\lambda=\mu$ and $\beta =1$ --- then

\bea
\overline{\Psi_{\mu}(q,p)} \,\Psi_{\mu}(q,p)= \frac{2}{\pi}\,e^{-2(q-x_C)^2-2(p-k_C)^2}\, 
\label{cs13}
\eea
which is also centered on the same values as the Wigner function.  Note however that  these two distributions 
(\ref{cs12}) and (\ref{cs13}) are  not equal, 
even in this case when both are positive. 
\vs\ni
Comparing (\ref{cs11}) with (\ref{cs12}) and (\ref{cs13}), we see also the effect of choosing a window that is {\em not} 
centred on the key features
of the Wigner function.

\section{Schr\"odinger's equation in phase space and evolution of amplitudes} 
\vs\ni
From (\ref{schro1}), corresponding to  an 
evolving state vector $|\psi(t)\rangle$ in $\Hil$, 
we have  for a time-dependent amplitude in phase space,  Schr\"odinger's 
equation in the form

\bea
i\,\partial_t \Psi(q,p,t)=H(q,p)\star \Psi(q,p,t)\,,
\label{schro2}
\eea
where $H=\W(\H)$. Supposing $H$ is not explicitly time-dependent, this equation integrates to give

\bea
\Psi(q,p,t)=U(t-t_0)\star\Psi(q,p,t_0)\,,\quad U(t)=e^{\star [-iH(q,p)t]}\,,
\label{evolution1}
\eea
where the star exponential is defined formally by

\bea
e^{\star [A]}= \W\left(e^{{\widehat A}}\right)=1+A+A\star A/2! + A\star A\star A/3!+\dots\,.
\label{star_exponential}
\eea
When $H$ is a polynomial in $q$ and $p$, a more explicit form for the time-evolution may be available. For example, 
if $H$ describes a non-relativistic particle and has the form

\bea
H(q,p)=\half p^2+V(q)\,,
\label{non_rel_Ham1}
\eea
with $V$ a polynomial in $q$, then from (\ref{star3}) we can rewrite (\ref{schro2}) as

\bea
i \,\partial_t \Psi(q,p,t)=H(\qBL,\pBL) \Psi(q,p,t)\,,
\label{schro3}
\eea
leading to

\bea
\Psi(q,p,t)=e^{-iH(\qBL,\,\pBL)(t-t_0)}\,\Psi(q,p,t_0)\,.
\label{evolution2}
\eea
\vs\ni
An efficient way to solve  (\ref{schro2}) or (\ref{schro3}) explicitly, when this is possible, is to solve Schr\"odinger's equation 
in the coordinate representation to get the wavefunction $\psi(x,t)$  and then use 
that in  (\ref{Psi_def2})   
to construct $\Psi(q,p,t)$. 
\vs\ni
For example, consider the initial-value problem for a free particle 

\bea
i\partial_t \psi(x,t)=-\half \partial^2\psi(x,t)/\partial x^2\,,\quad \psi(x,0)=\sqrt{\gamma}\,e^{- \gamma^2x^2/2}/\pi^{1/4}
\label{IV_problem1}
\eea  
where $\gamma$ is a positive constant, with solution 

\bea
\psi(x,t)=\sqrt{\gamma}\,e^{-\gamma^2x^2/2g(t)^2}/g(t)\pi^{1/4}\,,\quad g(t)=[1 +i\gamma^2t]^{1/2}\,,\quad t\geq 0\,.
\label{evolution3}
\eea
Here the branch of the complex square-root is chosen so that $g(0)=1$.  
With the Gaussian (\ref{gaussian5}) as window, (\ref{evolution3}) gives the corresponding  time-dependent phase space amplitude

\bea
\Psi(q,p,t)= [4\beta\gamma/\pi(\beta^2g(t)^2+\gamma^2)]^{1/2}\,
e^{\beta^2{\bar\lambda}({\bar\lambda}-\lambda)/4}\,
e^{-\beta^2{\bar\lambda}^2\gamma^2/2(\beta^2g(t)^2+\gamma^2)}
\mea\mea
\times e^{-2[\beta^2 \gamma^2 q^2+p^2 g(t)^2 +i(\beta^2 g(t)^2-\gamma^2 )qp-\beta^2\gamma^2
{\bar \lambda}q-i\beta^2{\bar \lambda}p g(t)^2]/(\beta^2 g(t)^2+\gamma^2)}\,,
\label{IV_problem3}
\eea
as the solution of 

\bea
i\,\partial_t \Psi(q,p,t)=\half \pBL^2 \Psi(q,p,t)\,,
\label{IV_problem4}
\eea
corresponding to the initial value

\bea
\Psi(q,p,0)= [4\beta\gamma/\pi(\beta^2+\gamma^2)]^{1/2}\,
e^{\beta^2{\bar\lambda}({\bar\lambda}-\lambda)/4}\,
e^{-\beta^2{\bar\lambda}^2\gamma^2/2(\beta^2+\gamma^2)}
\mea\mea
\times e^{-2[\beta^2 \gamma^2 q^2+p^2  +i(\beta^2 -\gamma^2 )qp-\beta^2\gamma^2
{\bar \lambda}q-i\beta^2{\bar \lambda}p ]/(\beta^2+\gamma^2)}\,,
\label{IV_problem5}
\eea
which may be compared with (\ref{cs10}) in the case $\gamma=1$, $\mu=0$.  
\vs\ni
If the eigenvalue problem

\bea
H(q,p)\star\Psi(q,p)=E\Psi(q,p)
\label{schro4}
\eea
can be solved to find a complete set of phase space eigenfunctions $\Psi_{E_n}(q,p)$ --- some will be generalized 
eigenfunctions if $\widehat{H}$ has a (partly) continuous spectrum --- then a more explicit form of solution to (\ref{schro2}) is

\bea
\Psi(q,p,t)=\Sigma_n c_n \,e^{-iE_n (t-t_0)}\,\Psi_{E_n}(q,p)\,,
\mea\mea 
\Psi(q,p,t_0)= \Sigma_n c_n \,\Psi_{E_n}(q,p)\,.\qquad
\label{schro5}
\eea
Here each sum must be extended to include an integral over the continuous spectrum, when appropriate. 
The eigenvalue problem (\ref{schro4}) is distinct from the `$\star$-genvalue' 
problem discussed in the context of the (one-state) Wigner function \cite{curtright,gosson3}, 
which can be expressed in terms of the solution of (\ref{schro4}) using (\ref{pseudoBorn}). 
\vs\ni
The expressions (\ref{schro5}) are the images under $\W$ of the coordinate space formulas

\bea
\psi(x,t)=\Sigma_n c_n \,e^{-iE_n (t-t_0)}\,\psi_n(x)\,,\quad \psi(x,t_0)=\Sigma_n c_n \,\psi_n(x)\,,
\mea\mea
{\rm where} \quad {\widehat H}\psi_n=E_n\psi_n\,.\qquad\qquad\qquad\qquad
\label{schro6}
\eea
Similarly, orthogonality of the coordinate space eigenfunctions and determination of the expansion coefficients, 

\bea
\int {\overline \psi_m(x)}\,\psi_n(x)\,dx =\delta_{mn}\,,\quad
c_n=\int {\overline \psi_n(x)}\,\psi(x,t_0)\,dx  
\label{schro6A}
\eea
have images 

\bea
\int {\overline \Psi_m(q,p)}\,\Psi_n(q,p)\,d\Gamma =\delta_{mn}\,,\quad
c_n=\int {\overline \Psi_n(q,p)}\,\Psi(q,p,t_0)\,d\Gamma\,.
\label{schro7}
\eea
If the $\psi_n$ form a complete orthonormal set in the coordinate space representation of Hilbert space, 
so the $\Psi_n$ form a complete orthonormal set in $\subspace$. 
\vs\ni 
The simple harmonic oscillator with $V(q)=\half  q^2$ provides the simplest illustration.  In this case, (\ref{star2}) shows that 
(\ref{schro4}) becomes

\bea
[\half (q^2+p^2)+\half i(q\partial_p -p\partial_q)-\textstyle{\frac{1}{8}}(\partial_q^2+\partial_p^2)]\Psi(q,p)=E\Psi(q,p)
\eea
which is easily solved using a phase space variant of the boson calculus.  Set 

\bea
{\check A}=[(q+\half i \partial_p)+i(p-\half i\partial_q)]/\sqrt{2}=\half \bar{w}+\partial_w\,,
\mea\mea
{\check A} ^{\dagger}=[(q+\half i \partial_p)-i(p-\half i\partial_q)]/\sqrt{2}=\half w -\partial_{\bar{w}}\,,
\label{phase_space_bosons}
\eea
where we have used checks to distinguish phase space operators, and again introduced $w=\sqrt{2}(q-ip)$ as in (\ref{cs7}).
\vs\ni
Solving ${\check A}\Psi_0=0$ 
for the phase space `vacuum state', normalizing it, and then setting $\Psi_n={\check A}^{\dagger n}\Psi_0/\sqrt{n!}$ for $n=1,\,2,\,\dots$, 
we find

\bea
\Psi_n(q,p)=\frac{1}{\sqrt{n!}} \sum_{m=0}^n C^{\,n}_m(-1)^{n-m}\,w^mF^{(n-m)}(\bar{w})\,e^{-\bar{w}w/2}\,,
\label{oscillator_state}
\eea
corresponding to the familiar eigenvalue $E_n=n+\half$, for $n=0,\,1,\,2,\,\dots$.  In (\ref{oscillator_state}), 
$F(\bar{w})$ and its derivatives are
determined by the choice of window and the normalization of $\Psi_0$, which requires

\bea
\int \overline{F(\bar{w})} F({\bar w})\,e^{-{\bar w}w}\,d^2w=1\,.
\label{F_normalization}
\eea  
The obvious choice in the present context, $\varphi_0(x)$ as in (\ref{gaussian5}) 
with $\beta=1$, $\lambda=0$, leads to $F=\sqrt{2/\pi}$, and (\ref{oscillator_state}) is then the Bargmann wavefunction 
for the $n$-th oscillator eigenstate, apart from the exponential factor.  
But again we emphasize that any convenient window function can be chosen, 
and every choice leads through (\ref{pseudoBorn}) to the same Wigner functions, which in this case are \cite{hillery}

\bea
W_n(q,p)=  (-1)^n\,L_n(2[q^2+p^2])\,e^{-(q^2+p^2)}/\pi \,,
\label{SHO_wigners}
\eea
where $L_n$ is the Laguerre polynomial \cite{abramowitz}.
Different choices of $\varphi_0$ lead to different phase space amplitudes $\Psi_n$ and hence to different 
distributions $|\Psi_n|^2$ but the same Wigner function (\ref{SHO_wigners}).  
For example, in the case $n=1$,  the choice (\ref{gaussian5}) with $\beta=1$, $\lambda=0$ 
gives from (\ref{oscillator_state}) 

\bea
\Psi_1(q,p)=(\sqrt{2/\pi}\,\,)\,w\,e^{-\bar{w} w/2}
\label{bargmann_approx1}
\eea
and hence
\bea
|\Psi_1(q,p)|^2 =(2/\pi)\,(\bar{w}w)\,e^{-\bar{w}w}\,,
\label{bargmann_approx2}
\eea
whereas the choice of a `square' window

\bea
\varphi_0(x)=\left\{\begin{array}{r@{\quad:\quad}l}
1/\sqrt{2a}&|x|<a\\
0&{\rm otherwise}
\end{array}\right.
\label{square_window1}
\eea 
leads to 
\bea
F(\bar{w})= (1/\pi^{1/4}\sqrt{a})\,e^{\bar{w}^2/2}\,\left[{\rm erf}(a/\sqrt{2}-\bar{w})+{\rm erf}(a/\sqrt{2}+\bar{w})\right]
\label{square_window2}
\eea
in (\ref{oscillator_state}) and from there to 

\bea
\Psi_1(q,p) = \qquad\qquad\qquad\qquad\qquad\qquad\qquad\qquad\qquad&&
\mea\mea
(1/\pi^{1/4}\sqrt{2a})\,e^{-\bar{w}w/2}\,\left\{(w-\bar{w})\,e^{\bar{w}^2/2}
\left[{\rm erf}(a/\sqrt{2}-\bar{w})+{\rm erf}(a/\sqrt{2}+\bar{w})\right]\right.&&
\mea\mea
\left. +4/\sqrt{\pi}\,e^{-(a^2+\bar{w}^2)/2}\,\sinh(\sqrt{2}a\bar{w})\right\}\,.& &
\label{square_window3}
\eea
Fig. 4 shows the Wigner function $W_1$ of (\ref{SHO_wigners}), and    $|\Psi_1|^2$ for 
the Gaussian window with $\beta=1$, $\lambda=0$ as in (\ref{bargmann_approx2}).  Fig. 5 shows
$|\Psi_1|^2$  for $\Psi_1$ as in (\ref{square_window3}) in the case of a 
square window with $a=1$.

\vs\ni
[Figs. 4 and 5 near here.]
\vs\ni

\section{Eigenstates of momentum and position in phase space}

\vs\ni
In view of (\ref{star3}), a (generalized) eigenstate of momentum in phase space is defined  by

\bea
\pBL\,\Psi_{k_0}(q,p)=k_0\Psi_{k_0}(q,p)\,,
\label{momentum_state1}
\eea
giving 

\bea
\Psi_{k_0}(q,p)= F_{k_0}(p)\,e^{-2i(p-k_0)q}\,,
\label{momentum_state2}
\eea
with $F_{k_0}(p)$ undetermined.  The form of $F_{k_0}$ depends on the choice of $|\varphi_0\rangle$.  
In fact, when the generalized momentum eigenfunction (plane wave)

\bea
\psi(x)=e^{ik_0x}/\sqrt{2\pi}
\label{config_momentum_state}
\eea
is inserted into (\ref{Psi_def2}), it is revealed that 

\bea
\Psi_{k_0}(q,p)=\,\sqrt{2/\pi}\,\,\overline{\tilde{\varphi}_0(2p-k_0)}\,e^{-2i(p-k_0)q}\,,
\label{momentum_state4}
\eea
where $\tilde{\varphi}_0 (k)=\langle k|\varphi_0\rangle$ is the function in momentum space that corresponds to  
($=$ Fourier transform of) $\varphi_0(x)$.  
It can be checked directly using the integral form of the star product 
\cite{neumann,groenewold,moyal,fairlie},

\bea
&\, & (A\star B)(q,p)
\mea\mea
&=&\frac{1}{\pi^2}\int_{\Gamma_2\times \Gamma_3} A(q+q_2,p+p_2)B(q+q_3,p+p_3)\,e^{2i(q_2 p_3-q_3 p_2)}
\mea\mea
&\,&\qquad\qquad \qquad\qquad\qquad\qquad\qquad\qquad\times \,d\,\Gamma_2\,d\,\Gamma_3\qquad\qquad\qquad
\label{star_integral}
\eea
that (\ref{pseudoBorn}) 
does indeed hold in this case whatever choice is made for $|\varphi_0\rangle$,  
in every case yielding the (singular) Wigner function 

\bea
W_{k_0}(q,p)=\delta(p-k_0)/2\pi\,. 
\label{momentum_wigner}
\eea 
\vs\ni
In a similar way, the generalized phase space position eigenfunction is defined by

\bea
\qBL\,\Psi_{x_0}(q,p)=x_0\Psi_{x_0}(q,p)
\label{position_state1}
\eea
and we find 

\bea
\Psi_{x_0}(q,p)=\sqrt{2/\pi}\,\,\overline{\varphi_0(2q-x_0)}\,e^{2ip(q-x_0)}\,,
\label{position_state2}
\eea 
corresponding to $W_{x_0}(q,p)=\delta(q-x_0)/2\pi$. Noting that $\Psi_{x_0}$ in  (\ref{position_state2}) 
is $\W(|x_0\rangle\langle\varphi_0|)/\sqrt{2\pi}$, 
we then see that the second of formulas (\ref{Psi_def2}) is simply the 
image under the Weyl-Wigner transform of the formula

\bea
|\psi\rangle\langle\varphi_0|=\int |u\rangle\langle u|\psi\rangle\langle\varphi_0|\,du=\int \psi(u)\,|u\rangle\langle\varphi_0|\,du\,.
\label{position_state3}
\eea 
\vs\ni
Formulas (\ref{momentum_state4}) and (\ref{position_state2}) show again the benefit of choosing a 
window state that is centered on a region of interest in phase space. For example 
if we choose a Gaussian window centered on coordinate and 
momentum values $x_W$ and $k_W$ as in (\ref{gaussian5}), 
(\ref{gaussian6}), (\ref{gaussian7}),
then for the phase space momentum eigenfunction (\ref{momentum_state4}) we get

\bea
\Psi_{k_0}(q,p)=(4/\pi^3 \beta^2)^{1/4}\,
e^{-(2p-k_0-k_W)^2/2\beta^2}
\mea\mea
\times e^{-i(2qp-2qk_0-2x_Wp+x_Wk_0+x_Wk_W/2)}\,.
\label{momentum_state5}
\eea
If the window state has $k_W=k_0$ and so is centred on the momentum value $k_0$ of interest, then

\bea
|\Psi_{k_0}(q,p)|^2 = (4/\pi^3 \beta^2)^{1/2}\,e^{-4(p\,- k_0)^2/\beta^2}\,,
\label{modPsisquared_1}
\eea
which is also centred on $p=k_0$ for each value of $q$, and may be compared with the Wigner function (\ref{momentum_wigner}).    

\vs\ni
Similarly, 
for the phase space position eigenstate (\ref{position_state2}) with a Gaussian window we get

\bea
\Psi_{x_0}(q,p)=(4\beta^2/\pi^3)^{1/4}\,
e^{-\beta^2(2q-x_0-x_W)^2/2}
\mea\mea
\times e^{i(2qp-2x_0p-2qk_W+x_0k_W+x_Wk_W/2)}\,,
\label{position_state4}
\eea
so that if the window state has $x_W=x_0$, then

\bea
|\Psi_{x_0}(q,p)|^2 = (2\beta^2/\pi^3 )^{1/2}\,e^{-2\beta^2(q\,- x_0)^2}\,,
\label{modPsisquared2}
\eea
centered on  $q=x_0$ for each value of $p$.  Fig. 6 shows the real part of $\Psi_{k_0}$
for $k_0=-2$, 
with
the parameter choices $k_W=k_0$, $x_W=4$ and $\beta=1$ in the Gaussian window (\ref{gaussian5}).  The imaginary part is similar in form.

\vs\ni
[Fig. 6 near here.]
\vs\ni

\section{Oscillator states as windows}
\vs\ni
An obvious generalization of the Gaussian window state with wavefunction $\varphi_0(x)$ as in (\ref{gaussian5}), 
is  the $n$-th excited state of the oscillator, for some nonnegative integer $n$,  
more precisely, the state obtained by applying a suitable `creation operator' $n$ times to the 
Gaussian `ground state', to give the normalized wavefunction
 
\bea
{\check \varphi}_0(x)&=&(1/\sqrt{n!})\,[(\beta (x-{\bar \lambda}) -\partial_x/\beta)/\sqrt{2}]^n\varphi_0(x)\,.
\label{osc_window1}
\eea
Windows of this type have been considered recently \cite{abreu} in the context 
of the Gabor transform; just as the Gaussian choice 
leads to the (generalized) Bargmann transform as noted above, 
so these oscillator windows lead to further generalizations of the Bargmann transform and 
Bargmann representation.

\vs\ni   
It follows at once from (\ref{osc_window1}) that if $\Psi (q,p)$ is the 
amplitude corresponding to a system state $|\psi\rangle$
and window state $|\varphi_0\rangle$, then the amplitude corresponding to the state $|\psi\rangle$ 
and window state $|{\check \varphi}_0\rangle$
is 

\bea
{\check \Psi}(q,p)=(1/\sqrt{n!})\,[(\beta (\qBR -\lambda)+i\pBR/\beta)/\sqrt{2}]^n\,\Psi(q,p)\,,
\label{osc_window2}
\eea
with $\qBR$, $\pBR$ as in (\ref{star3}).

\vs\ni
For example, the coherent state $\psi_{\mu}$ defined by (\ref{cs5}) leads to the amplitude $\Psi_{\mu}$ as in (\ref{cs10}) 
when the Gaussian window (\ref{gaussian5}) is chosen, whereas the choice of the $n=1$ (`Mexican Hat') excited state 
in (\ref{osc_window1}) as window leads through (\ref{osc_window2}) to the amplitude

\bea
{\check \Psi}_{\mu}(q,p)=\qquad\qquad\qquad\qquad\qquad\qquad\qquad\qquad\qquad\qquad
\mea\mea
\sqrt{2}\beta [(2q-x_W-x_C)+i(2p-k_W-k_C)]\,\Psi_{\mu}(q,p)/(\beta^2+1)\,,
\label{osc_window3}
\eea
and hence from (\ref{cs11}) to the non-negative expression

\bea
\overline{{\check \Psi}_{\mu}(q,p)}\,{\check \Psi}_{\mu}(q,p)
\qquad\qquad\qquad\qquad\qquad
\qquad\qquad\qquad\qquad\qquad\qquad
\mea\mea
=\frac{32\beta^3}{\pi(\beta^2 +1)^3}[(q-\eta)^2+(p-\zeta)^2]\,e^{-4[\beta^2(q-\eta)^2+(p-\zeta)^2]/(\beta^2+1)}\,,
\label{osc_window4}
\eea
with $\eta$, $\zeta$ as in (\ref{cs11}).  
This expression should be compared with $|\Psi_{\mu}(q,p)|^2$ as in (\ref{cs11}), and with $W_{\mu}(q,p)$ as in (\ref{cs12}).

\section{Concluding remarks}

\vs\ni
The introduction of phase space amplitudes extends the phase space formulation of 
quantum mechanics, and may be considered 
to complete that formulation by providing images of not only density operators but also state vectors.  

\vs\ni
The degree of arbitrariness in the definition of the amplitude chosen to represent 
a given quantum state vector reflects the freedom to choose any normalized state as window state, 
and may seem surprising at first, notwithstanding our final remarks in Section 2.  
Further study is needed to see how to optimize the choice of window for a given quantum system in
a given state, for purposes of computation and visualization.  
Selecting a Gaussian window 
with adjustable 
width and location in coordinate space and momentum space 
is simple and natural, and often leads to analytic expressions 
for amplitudes representing simple quantum states, 
as we have seen.  This has been useful for our purpose here, which was to illustrate basic ideas,  
but it may not always be optimal.  A natural choice of window for an evolving 
quantum system with a well-defined ground state and excited states might 
be its ground state, for example. 

\vs\ni
Some of the analysis of amplitudes corresponding to quantum states at a given instant,  mirrors studies 
of signals in the time-frequency domain done over many years using Gabor's  
windowed Fourier transform \cite{cohen2,ascensi,abreu}.
A feature of the situation in quantum mechanics that is absent in the case of signal 
processing arises from the very different 
role that the time variable plays in the two cases.  In the quantum case, the 
time is not one of the phase space variables, 
and the description of quantum phase space amplitudes evolving in time is a 
feature absent in the case of signals.  
Given that the window state once chosen is fixed, and does not evolve in time, 
the choice of an optimal window is 
therefore a more complicated problem in the quantum case. 

\vs\ni
Other questions that seem worthy of further study in terms of phase space amplitudes 
include the formulation of the uncertainty principle, the description of 
interference effects, and the description of quantum symmetries.  
We hope to return to some of these questions.

\vs\ni  
{\it Acknowledgements:}    Thanks to D.R. Stevens for a stimulating conversation, 
and to referees for helpful suggestions.

\newpage
\begin{figure}[t]
\centering
\includegraphics[width=13cm]{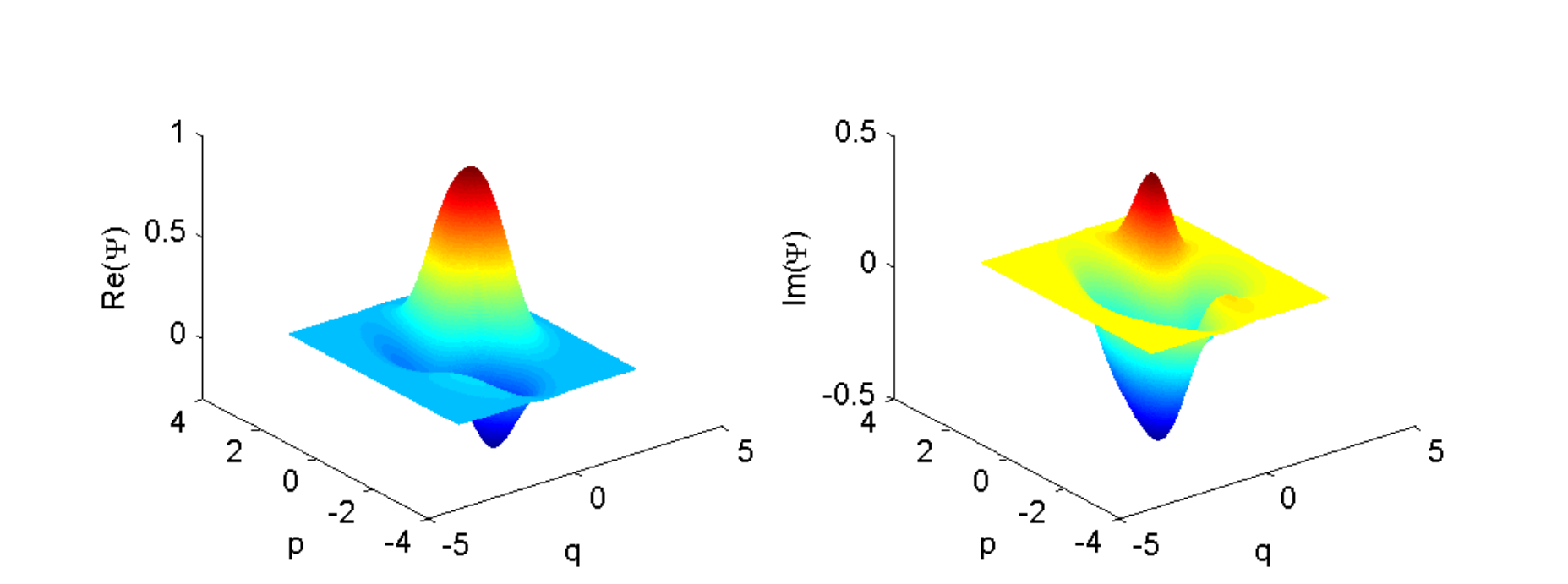}
\caption{The real and imaginary parts of the phase space amplitude (\ref{test3}) with $\beta=1$.}
\end{figure}

\vs\ni
\quad

\newpage
\quad

\begin{figure}[t]
\centering
\includegraphics[width=13cm]{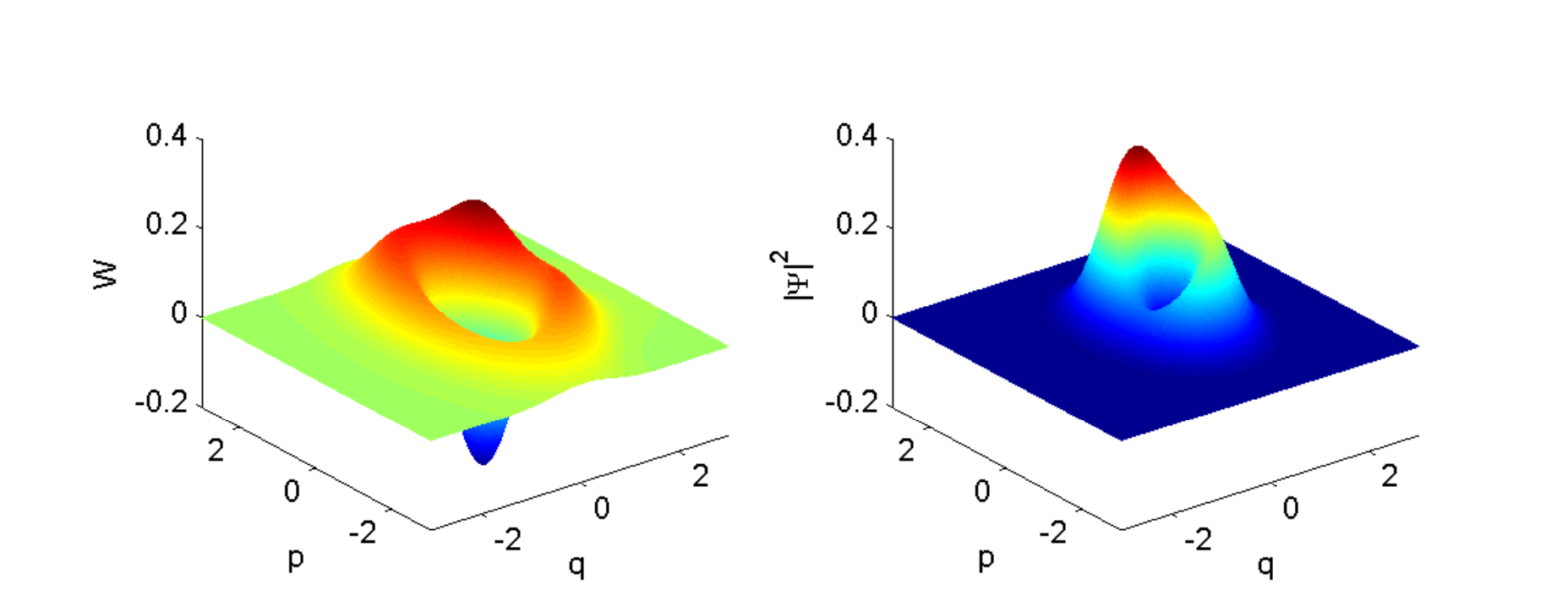}
\caption{Left to right: the Wigner function (\ref{test2}) and $|\Psi|^2$, 
for $\Psi$ as in (\ref{test3}) and $\beta=1$.}
\end{figure}

\begin{figure}[t]
\centering
\includegraphics[width=13cm]{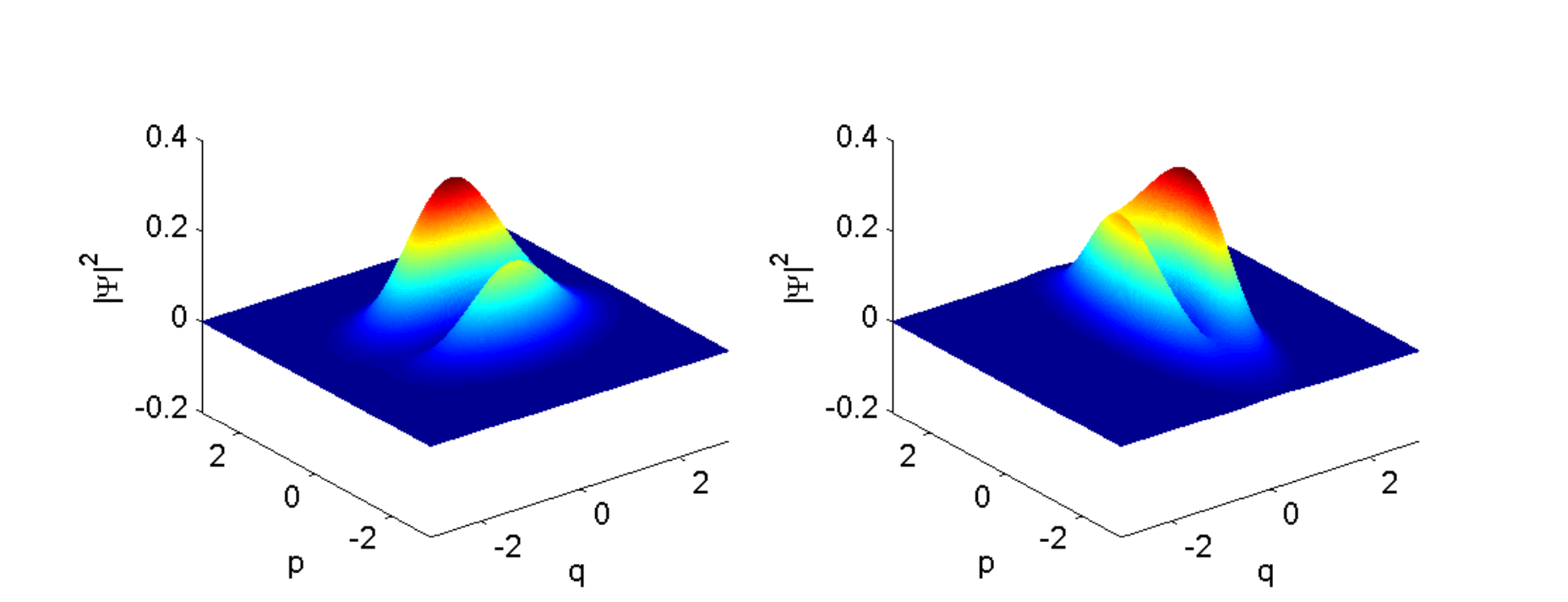}
\caption{Left to right: $|\Psi|^2$, for $\Psi$ as in (\ref{test3}) and $\beta=0.5$, $2$.}
\end{figure}

\newpage
\quad

\begin{figure}[t]
\centering
\includegraphics[width=13cm]{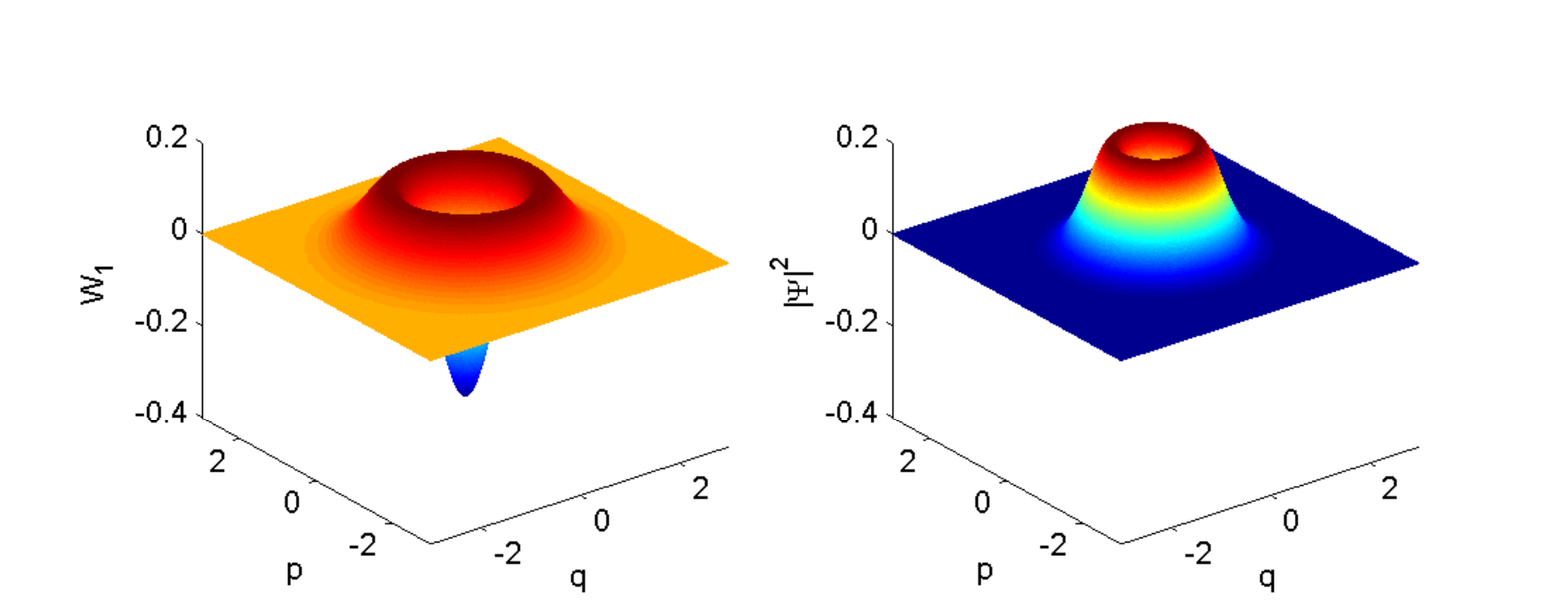}
\caption{The Wigner function $W_1$ of (\ref{SHO_wigners}) and $|\Psi_1|^2$ 
as in (\ref{bargmann_approx2}), for a Gaussian window
having $\beta=1$ and $\lambda=0$.}
\end{figure}

\newpage
\quad

\begin{figure}[t]
\centering
\includegraphics[width=10cm]{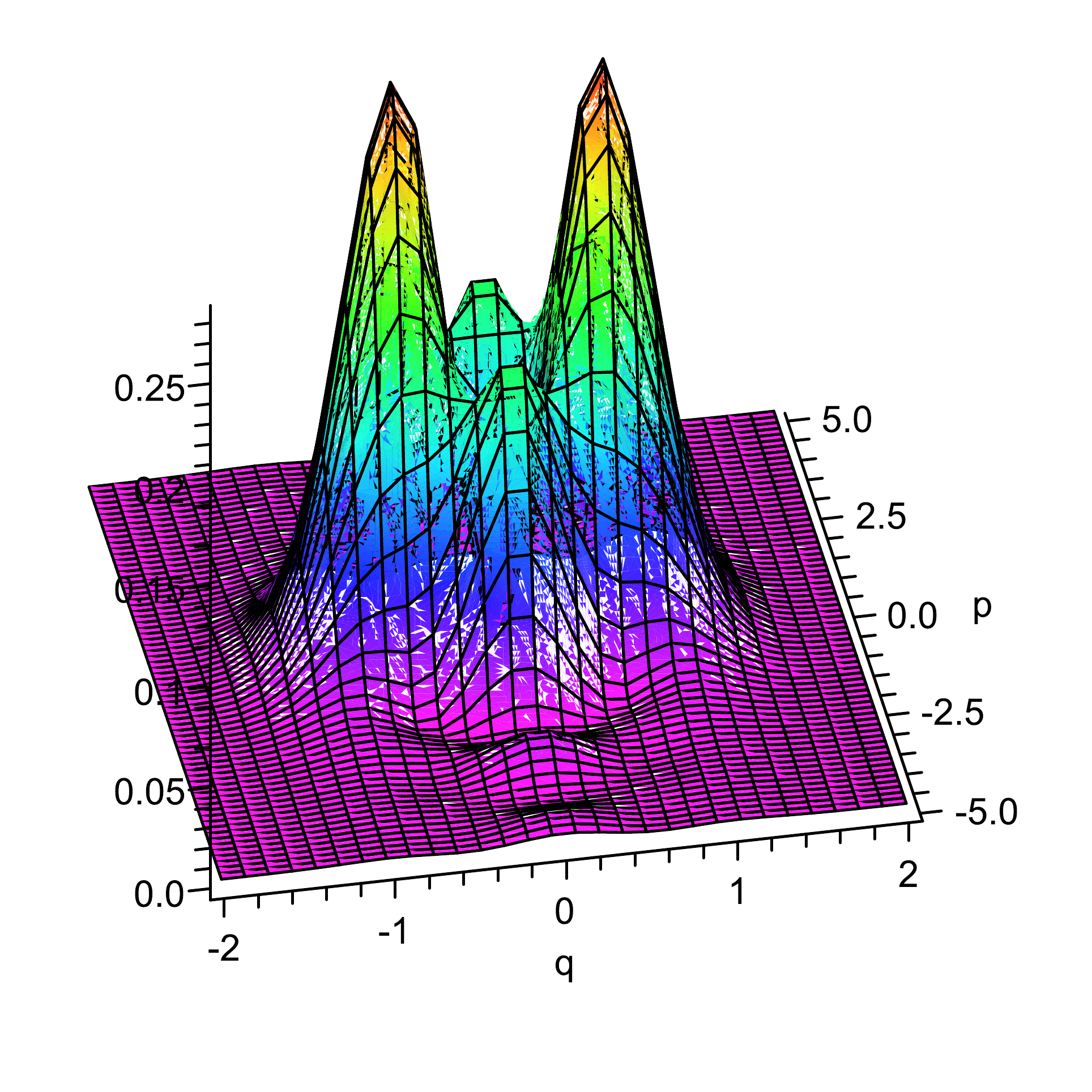}
\caption{The quantity $|\Psi_1|^2$ for $\Psi_1$ as in (\ref{square_window3}), 
for a square window having $a=1$.}
\end{figure}

\newpage
\quad

\begin{figure}[t]
\centering
\includegraphics[width=13cm]{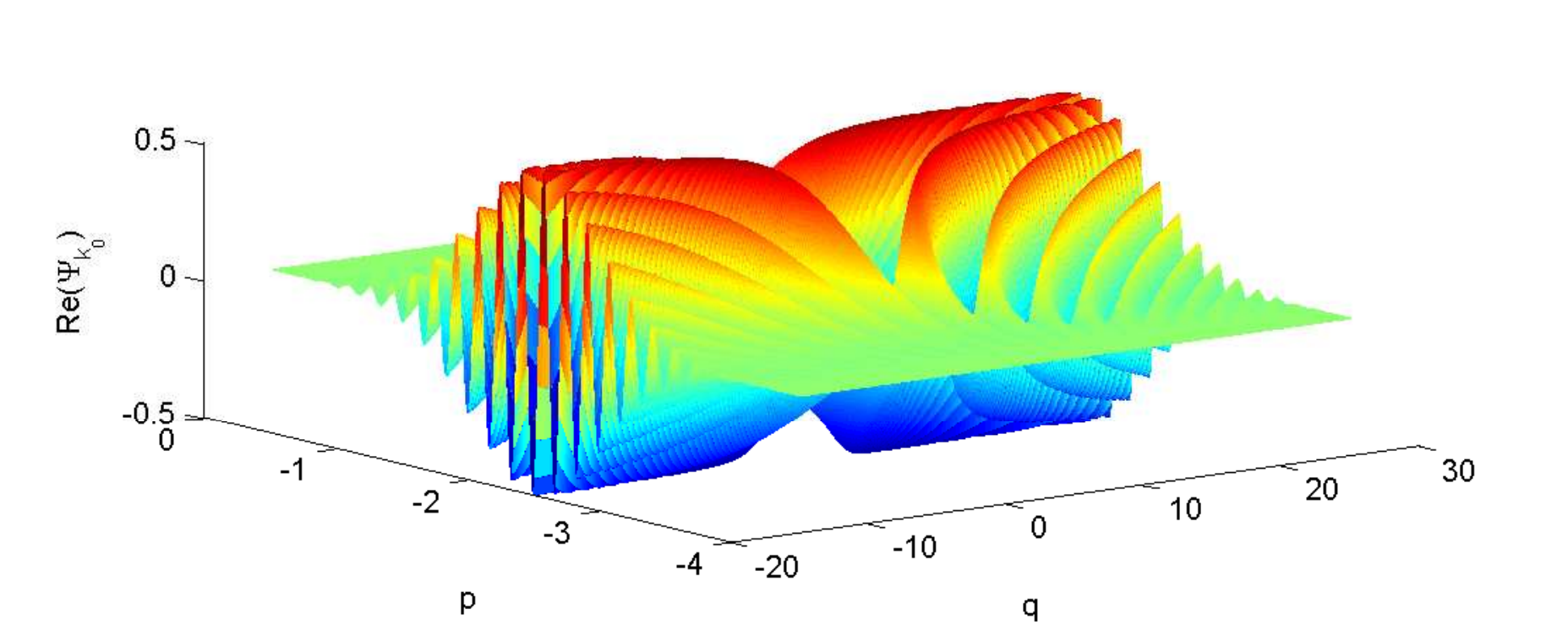}
\caption{${\rm Re}(\Psi_{k_0=-2})$ for the amplitude $\Psi_{k_0}$ as in (\ref{momentum_state5}), 
with a Gaussian window having $x_W=4$, $k_W=-2$ and $\beta=1$.}
\end{figure}

\end{document}